\documentclass[acmtocl]{acmtrans2m}
\usepackage{epsfig,color,amssymb}

\newtheorem{theorem}{Theorem}[section]

\newtheorem{lemma}[theorem]{Lemma}
\newdef{definition}[theorem]{Definition}
\newdef{remark}[theorem]{Remark}

\newcommand{\fsgd}{F_\sigma \cap G_\delta}
\newcommand{\nats}{\mathbb{N}}
\newcommand{\reals}{\mathbb{R}}
\newcommand{\ints}{\mathbb{Z}}

\newcommand{\infi}{\mbox{\it inf\/}}
\newcommand{\sdiff}{\, \setminus \,}

\title{An Effective Decision Procedure for Linear Arithmetic
with Integer and Real Variables$~^{\!\star}$}

\author{BERNARD BOIGELOT, S{\'E}BASTIEN JODOGNE$~^{\!\dagger}$, and
	PIERRE WOLPER \\
	Universit\'e de Li\`ege \\ 
	Institut Montefiore, B28 \\ 
	4000 Li\`ege, Belgium}

\begin{abstract}
This paper considers finite-automata based algorithms for handling
linear arithmetic with both real and integer variables. Previous work
has shown that this theory can be dealt with by using finite automata
on infinite words, but this involves some difficult and delicate to
implement algorithms. The contribution of this paper is to show,
using topological arguments, that only a restricted class of automata
on infinite words are necessary for handling real and integer linear
arithmetic. This allows the use of substantially simpler algorithms,
which have been successfully implemented. 
\end{abstract}




\begin{document}

\begin{bottomstuff}
Authors' e-mail~: {\tt \{boigelot,jodogne,pw\}@montefiore.ulg.ac.be}\\
Authors' website~:
	{\tt http://www.montefiore.ulg.ac.be/}$\sim$%
	{\tt \{boigelot,jodogne,pw\}/}
\medskip

$~^{\!\star}$ This work was partially funded
by a grant of the ``Communaut\'e fran\c{c}aise de Belgique -
Direction de la recherche scientifique - Actions de recherche
concert\'ees'' and by the European IST-FET project
{\sc Advance} (IST-1999-29082).\\
A preliminary version of this paper appeared as~\cite{BJW01}.
\medskip

$~^{\!\dagger}$ Research Fellow (``Aspirant'')
for the National Fund for Scientific Research (Belgium).

\medskip
\end{bottomstuff}

\maketitle

%Introduction
\section{Introduction}

Among the techniques used to develop algorithms for deciding or
checking logical formulas, finite automata have played an important
role in a variety of cases. Classical examples are the use of
infinite-word finite automata by B\"uchi~\cite{Buc62} for obtaining
decision procedures for the first and second-order monadic theories
of one successor, as well as the use of tree automata by
Rabin~\cite{Rab69} for deciding the second-order monadic theory of $n$
successors. More recent examples are the use of automata for
obtaining decision and model-checking procedures for temporal and
modal logics~\cite{VW86b,VW86,VW94,KVW00}. In this last setting,
automata-based procedures have the advantage of moving the
combinatorial aspects of the procedures to the context of automata,
which are simple graph-like structures well adapted to algorithmic
developments. This separation of concerns between the logical and
the algorithmic has been quite fruitful for instance in the
implementation of model checkers for linear-time temporal
logic~\cite{CVWY90,Hol97}.

As already noticed by B\"uchi~\cite{Buc62,Buc60}, automata-based
approaches are not limited to sequential and modal logics, but can
also be used for Presburger arithmetic. To achieve this, one adopts the
usual encoding of integers in a base $r \geq 2$, thus representing an
integer as a word over the alphabet $\{0, \ldots, r-1\}$. By
extension, $n$-component integer vectors are represented by words
over the alphabet $\{0, \ldots, r-1\}^n$ and a finite automaton
operating over this alphabet represents a set of integer
vectors. Given that addition and order are easily represented by
finite automata and that these automata are closed under Boolean
operations as well as projection, one easily obtains a decision
procedure for Presburger arithmetic. This idea was first explored at
the theoretical level, yielding for instance the very nice result
that base-independent finite-automaton representable sets are exactly
the Presburger sets~\cite{Cob69,Sem77,BHMV94}. Later, it has been proposed
as a practical means of deciding and manipulating Presburger
formulas~\cite{BC96,Boi98,SKR98,WB00}. The intuition behind this applied
use of automata for Presburger arithmetic is that finite automata play with
respect to Presburger arithmetic a role similar to the one of Binary
Decision Diagrams (BDD) with respect to Boolean logic. These ideas
have been implemented in the LASH tool~\cite{Boi99}, which has been
used successfully in the context of verifying systems with unbounded
integer variables.

It almost immediately comes to mind that if a finite word over the
alphabet $\{0, \ldots, r-1\}$ can represent an integer, an infinite
word over the same alphabet extended with a fractional part separator
(the usual dot) can represent a real number. Finite automata on
infinite words can thus represent sets of real vectors, and serve
as a means of obtaining a decision procedure for real additive
arithmetic. Furthermore, since numbers with fractional parts equal to
zero can easily be recognized by automata, the same technique can be
used to obtain a decision procedure for a theory combining the integers
and the reals. This was not previously handled by any tool, but can be
of practical use, for instance in the verification of timed systems
using integer variables~\cite{BBR97}. However, turning this into an
effective implemented system is not as easy as it might first
seem. Indeed, projecting and complementing finite automata on
infinite words is significantly more difficult than for automata on
finite words. Projection yields nondeterministic automata and
complementing or determinizing infinite-word automata is a
notoriously difficult problem. A number of algorithms have been
proposed for this~\cite{Buc62,SVW87,Saf88,KV97}, but even though their
theoretical complexity remains simply exponential as in the finite-word
case, it moves up from $2^{O(n)}$ to $2^{O(n \log n)}$ and none of
the proposed algorithms are as easy to implement and fine-tune 
as the simple Rabin-Scott subset construction used in the finite-word case.

However, it is intuitively surprising that handling reals is so much
more difficult than handling integers, especially in light of the fact
that the usual polyhedra-based approach to handling arithmetic is both
of lower complexity and easier to implement for the reals than for the
integers~\cite{FR79}. One would expect that handling reals with
automata should be no more difficult than handling
integers\footnote{Note that one cannot expect reals to be easier to
handle with automata than integers since, by nature, this
representation includes explicit information about the existence of
integer values satisfying the represented formula.}. The conclusion
that comes out of these observations is that infinite-word automata
constructed from linear arithmetic formulas must have a special
structure that makes them easier to manipulate than general automata
on infinite words. That this special structure exists and that it can
exploited to obtain simpler algorithms is precisely the subject of
this paper.

As a starting point, let us look at the topological characterization
of the sets definable by linear arithmetic formulas. Let us first
consider a formula involving solely real variables. If the formula is
quantifier free, it is a Boolean combination of linear constraints and
thus defines a set which is a finite Boolean combination of open and
closed sets. Now, since real linear arithmetic admits quantifier
elimination, the same property also holds for quantified
formulas. Then, looking at classes of automata on infinite words, one
notices that the most restricted one that can accept Boolean
combinations of open and closed sets is the class of deterministic
weak automata~\cite{SW74,Stai83}. These accept all $\omega$-regular
sets in the Borel class $F_\sigma \cap G_\delta$ and hence also finite
Boolean combinations of open and closed sets. So, with some care about
moving from the topology on vectors to the topology on their encoding
as words, one can conclude that the sets representable by arithmetic
formulas involving only real variables can always be accepted by
deterministic weak automata on infinite words. 
If integers are also involved in the formula, a similar argument can
be used, invoking a recently published quantifier elimination result
for the combined theory~\cite{Wei99}. However, initially unaware of
this result, we developed a different argument to prove that sets
definable by quantified linear arithmetic formulas involving both
real and integer variables are within $F_\sigma \cap G_\delta$ and
thus are representable by weak deterministic automata. This proof
relies on separating the integer and fractional parts of variables
and on topological properties of $F_\sigma \cap G_\delta$. It has the
advantage of being much more direct than a proof relying on a
quantifier elimination result.

The problematic part of the operations on automata used for deciding
a first-order theory is the sequence of projections and
complementations needed to eliminate a string of quantifiers
alternating between existential and universal ones. The second result
of this paper shows that for sets defined in linear arithmetic this
can be done with constructions that are simple adaptations of the
ones used for automata on finite words. Indeed, deterministic weak
automata can be viewed as either B\"uchi or co-B\"uchi automata.  The
interesting fact is that co-B\"uchi automata can be determinized by
the ``breakpoint'' construction~\cite{MH84,KV97}, which basically
amounts to a product of subset constructions. Thus, one has a simple
construction to project and determinize a weak automaton, yielding a
deterministic co-B\"uchi automaton, which is easily complemented into
a deterministic B\"uchi automaton.  In the general case, another
round of projection will lead to a nondeterministic B\"uchi
automaton, for which a general determinization procedure has to be
used. However, we have the result that for automata obtained from
linear arithmetic formulas, the represented sets stay within those
accepted by deterministic weak automata. We prove that this implies
that the automata obtained after determinization will always be weak.

Note that this cannot be directly concluded from the fact that the
represented sets stay within those representable by deterministic
weak automata. Indeed, even though the represented sets can be
accepted by deterministic weak automata, the automata that are
obtained by the determinization procedure might not have this
form. Fortunately, we can prove that this is impossible. For this, we
go back to the link between automata and the topology of the sets of
infinite words they accept. The argument is that $\omega$-regular
sets in $F_\sigma \cap G_\delta$ have a topological property that
forces the automata accepting them to be inherently weak, i.e. not
to have strongly connected components containing both accepting and
non accepting cycles.

Finally, an important additional benefit of working with weak
deterministic automata is that they admit a canonical minimal normal
form that can be obtained efficiently~\cite{MS97,Loe01}. This brings
us even closer to the situation of working with finite-work automata,
and is a property that is not available when working either with
general infinite-word automata, or with formulas as done
in~\cite{Wei99}.

As a consequence of our results, we obtain a much simplified decision
procedure for the theory combining integer and real linear arithmetic.
The fact that this theory is decidable using automata-based methods
was known~\cite{BBR97}, but the results of this paper make it
possible to implement a tool that can handle it effectively.

%Automata-Theoretic background
\section{Automata-Theoretic and Topological Background}
\label{sec-background}

In this section we recall some automata-theoretic and topological
concepts that are used in the paper. 

\subsection{Automata on Infinite Words}
\label{sec-background-auto}

An infinite word (or $\omega$-word) $w$ over an alphabet $\Sigma$ is
a mapping $w : \nats
\mapsto \Sigma$ from the natural numbers to $\Sigma$. A B\"uchi
automaton on infinite words is a five-tuple
$A=(Q,\Sigma,\delta,q_0,F)$, where
\begin{itemize}
\item
$Q$ is a finite set of states;
\item
$\Sigma$ is the input alphabet;
\item
$\delta$ is the transition function and is of the form $\delta : Q
\times \Sigma \mapsto 2^Q$ if the automaton is nondeterministic
and of the form  $\delta : Q \times \Sigma \mapsto Q$ if the
automaton is deterministic;
\item
$q_0$ is the initial state;
\item
$F$ is a set of accepting states.
\end{itemize}

A run $\pi$ of a B\"uchi automaton $A=(Q,\Sigma,\delta,q_0,F)$ on an
$\omega$-word $w$ is a mapping $\pi : \nats \mapsto Q$ that satisfies the
following conditions~:
\begin{itemize}
\item
$\pi(0)= q_0$, i.e. the run starts in the initial state;
\item
for all $i \geq 0$, $\pi(i+1) \in \delta(\pi(i), w(i))$
(nondeterministic automata) or  $\pi(i+1) =  \delta(\pi(i), w(i))$
(deterministic automata), i.e. the run respects the transition
function.
\end{itemize}

Let $\infi (\pi)$ be the set of states that occur infinitely often in a
run $\pi$. A run $\pi$ is said to be accepting if $\infi (\pi) \cap F
\neq \emptyset$. An $\omega$-word $w$ is accepted by a B\"uchi
automaton if that automaton has some accepting run on $w$. The
language $L_\omega(A)$ of infinite words defined by a B\"uchi
automaton $A$ is the set of $\omega$-words it accepts. The
$\omega-$regular languages are defined as the languages of
infinite words that can be accepted by a nondeterministic B\"uchi
automaton.

A co-B\"uchi automaton is defined exactly as a B\"uchi automaton
except that its accepting runs are those for which $\infi (\pi) \cap F
= \emptyset$. 

We will also use the notion of {\em weak} automata~\cite{MSS86}.  For
a B\"uchi automaton $A=(Q,\Sigma, \delta, q_0, F)$ to be weak,
there has to be a partition of its state set $Q$ into disjoint
subsets $Q_1,\ldots, Q_m$ such that 
\begin{itemize}
\item
for each of the $Q_i$ either $Q_i
\subseteq F$ or $Q_i \cap F = \emptyset$, and
\item
there is a partial order $\leq$ on the sets $Q_1,\ldots, Q_m$ such
that for every $q \in Q_i$ and $q' \in Q_j$ for which, for some $a
\in \Sigma$, $q' \in \delta(q,a)$ ($q' = \delta(q,a)$ in the
deterministic case), $Q_j \leq Q_i$.
\end{itemize}

For more details, a survey of automata on infinite words can be found
in~\cite{Tho90}.

\subsection{Topology}
\label{topology}

\begin{figure}[ht]
\begin{center}
\input{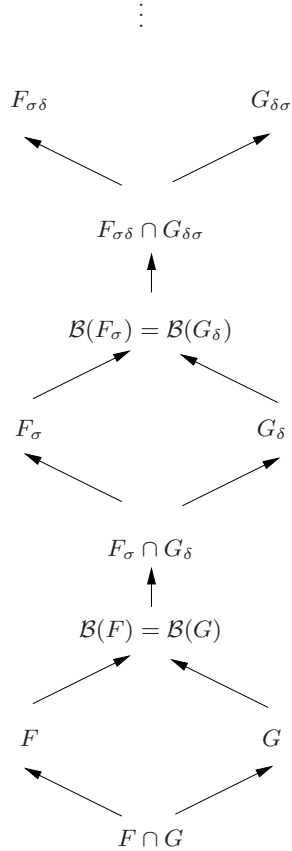}
\end{center}
\caption{The first few levels of the Borel hierarchy in a
metric topology.~\label{fig-borel}}
\end{figure}

Given a set $S$, a distance $d(x,y)$ defined on this set induces a
metric  topology on subsets of $S$. A neighborhood $N_{\varepsilon}(x)$
of a point $x \in S$ with respect to $\varepsilon \in \reals^+$ is
the set $N_{\varepsilon}(x) =\{y \mid d(x,y) < \varepsilon\}$.
A set $C \subseteq S$ is said to be open if for all 
$x \in C$, there exists $\varepsilon > 0$ such that the neighborhood
$N_{\varepsilon}(x)$ is contained in $C$. A closed set is a set whose
complement with respect to $S$ is open. We will be referring to the
first few levels of the Borel hierarchy which are shown in
Figure~\ref{fig-borel}. The notations used are the following~:
\begin{itemize}
\item
$F$ are the closed sets,
\item
$G$ are the open sets,
\item
$F_\sigma$ is the class of countable unions of closed sets,
\item
$G_\delta$ is the class of countable intersections of open sets,
\item
$F_{\sigma\delta}$ is the class of countable intersections of $F_\sigma$
sets,
\item
$G_{\delta\sigma}$ is the class of countable unions of $G_\delta$ sets,
\item
${\cal B} (X)$ represents the finite Boolean combinations of sets in
$X$.
\end{itemize}
An arrow between classes indicates proper inclusion.

%Topological Characterization of Arithmetic Sets
\section{Topological Characterization of Arithmetic Sets}
\label{sec-arith}

We consider the theory $\langle {\reals}, {\ints}, +, \leq
\nolinebreak \rangle$, where $+$ represents the predicate $x+y =z$.
Since any linear equality or order constraint can be encoded into
this theory, we refer to it as additive or linear arithmetic over
the reals and integers. It is the extension of Presburger arithmetic
that includes both real and integer variables. We provide the space
$\reals^n$ ($n\geq 0$) with the classical Euclidean distance between
vectors defined by
\[
d(\vec{x},\vec{y}) = \left( \sum_{i = 1}^n (x_i - y_i)^2\right)^{1/2}.
\]
The topology induced by this metric will be referred to as the
{\em natural topology\/} of $\reals^n$.

In this section, we prove that the sets representable in the additive
linear arithmetic over the reals and integers belong to the topological
class $\fsgd$. This result is formalized by the following
theorem.

\pagebreak
\begin{theorem}
Let $S \subseteq {\reals}^n$, with $n > 0$, be a set defined in 
the theory $\langle {\reals}, {\ints}, +,$ $\leq\nolinebreak
\rangle$. This set belongs to the class $F_\sigma \cap G_\delta$
of the natural topology of ${\reals}^n$.
\end{theorem}

\begin{proof}
Since $\langle {\reals}, {\ints}, +, \leq \nolinebreak \rangle$ is closed
under negation, it is actually sufficient to show that each formula
of this theory defines a set that belongs to $F_\sigma$, i.e., a set
that can be expressed as a countable union of closed sets.

Let $\varphi$ be a formula of $\langle {\reals}, {\ints}, +, \leq
\nolinebreak \rangle$. To simplify our argument, we will assume that
all free variables of $\varphi$ are reals. This can be done without
loss of generality since quantified variables can range over both
$\reals$ and $\ints$. We introduce $u < v$ as a shorthand for
$u \leq v \wedge \neg(u = v)$.

The first step of our proof consists of modifying $\varphi$ in the
following way. We replace each variable $x$ that appears in $\varphi$
by two variables $x_I$ and $x_F$ representing respectively the integer
and the fractional part of $x$. Formally, this operation replaces each
occurrence in $\varphi$ of a free variable $x$ by the sum $x_I + x_F$
while adding to $\varphi$ the constraints $0 \leq x_F$ and $x_F < 1$,
and transforms the quantified variables of $\varphi$ according to the
following rules~:
\begin{eqnarray*}
(\exists x\in {\reals})\phi & {\,\longrightarrow\,} & (\exists x_I\in
{\ints}) (\exists x_F\in {\reals}) (0 \leq x_F \wedge x_F < 1 \wedge
\phi[x / x_I + x_F])\\
 (\forall x\in {\reals})\phi &
{\,\longrightarrow\,} & (\forall x_I\in {\ints}) (\forall x_F\in 
{\reals}) (x_F < 0 \vee 1 \leq x_F \vee \phi[x / x_I + x_F])\\
(Q x\in {\ints})\phi & {\,\longrightarrow\,} & (Q x_I\in {\ints})\phi[x / x_I],
\end{eqnarray*}
where $Q \in \{\exists, \forall\}$, $\phi$ is a subformula, and
$\phi[x/y]$ denotes the result of replacing by $y$ each occurrence of
$x$ in $\phi$.  The transformation has no influence on the set
represented by $\varphi$, except that the integer and fractional parts
of each value are now represented by two distinct variables.

Now, the atomic formulas of $\varphi$ are of the form $p = q + r$, $p
= q$ or $p \leq q$, where $p, q$ and $r$ are either integer variables,
sums of an integer and of a fractional variable, or integer
constants. The second step consists of expanding these atomic formulas
so as to send into distinct atoms the occurrences of the integer and
of the fractional variables. This is easily done with the help of
simple arithmetic rules, for the truth value of the atomic formulas
that involve both types of variables has only to be preserved for
values of the fractional variables that belong to the interval $[0, 1)$.
The set of expansion rules\footnote{In these rules,
the expression $p = q + r + s$ is introduced as a shorthand for
$(\exists u\in \reals)(u = q + r\,\wedge\, p = u + s)$.} (up to
commutability of members and terms) is given in Figure~\ref{fig-rules}.

\begin{figure}
\small
$$\begin{array}{rcl}
x_I = (y_I+y_F) & \longrightarrow & x_I=y_I \wedge y_F=0 \\
(x_I+x_F)=(y_I+y_F) & \longrightarrow & x_I=y_I \wedge x_F=y_F \\
x_I = y_I + (z_I+z_F) & \longrightarrow & x_I=y_I+z_I \wedge z_F=0 \\
x_I = (y_I+y_F) + (z_I+z_F) & \longrightarrow &
	(x_I=y_I+z_I \wedge y_F+z_F=0)~\vee \\
	&&(x_I=y_I+z_I+1 \wedge y_F+z_F=1)\\
(x_I+x_F) = y_I+z_I & \longrightarrow & x_I=y_I+z_I \wedge x_F=0 \\
(x_I+x_F) = y_I+(z_I+z_F) & \longrightarrow & x_I=y_I+z_I \wedge x_F=z_F \\
(x_I+x_F) = (y_I+y_F)+(z_I+z_F) & \longrightarrow &
	(x_I=y_I+z_I \wedge x_F=y_F+z_F)~\vee \\
	&&(x_I=y_I+z_I+1 \wedge x_F=y_F+z_F-1)\\
x_I\leq (y_I+y_F) & \longrightarrow & x_I\leq y_I \\
(x_I+x_F) \leq y_I & \longrightarrow & x_I<y_I \vee (x_I=y_I \wedge x_F=0) \\
(x_I+x_F) \leq (y_I+z_I) & \longrightarrow &
	x_I<y_I \vee (x_I=y_I \wedge x_F\leq y_F)
\end{array}$$
\caption{Expansion rules.}
\label{fig-rules}
\end{figure}

After the transformation, each atomic formula of $\varphi$ is either
a formula $\phi_I$ involving only integer variables or a formula
$\phi_F$ over fractional variables. We now distribute existential
(resp. universal) quantifiers over disjunctions (resp. conjunctions),
after rewriting their argument into disjunctive (resp. conjunctive)
normal form, and then apply the simplification rules
\begin{eqnarray*}
(Q x_I \in {\ints})(\phi_I \,\alpha\, \phi_F) & \,\longrightarrow\, &
(Q x_I \in {\ints})(\phi_I) ~\alpha~ \phi_F \\
(Q x_F \in {\reals})(\phi_I \,\alpha\, \phi_F) & \,\longrightarrow\, &
\phi_I ~\alpha~ (Q x_F \in {\reals})(\phi_F),
\end{eqnarray*}
where $Q \in \{ \exists, \forall \}$ and $\alpha \in \{ \vee, \wedge
\}$.

Repeating this operation, we eventually get a formula $\varphi'$
equivalent to $\varphi$ that takes the form of a finite Boolean
combination $${\cal B}(\phi^{(1)}_I,
\phi^{(2)}_I, \ldots, \phi^{(m)}_I, \phi^{(1)}_F,
\phi^{(2)}_F,\ldots,\phi^{(m')}_F)$$ of subformulas 
$\phi^{(i)}_I$ and $\phi^{(i)}_F$ that involve respectively only
integer and fractional variables.

Let $x^{(1)}_I, x^{(2)}_I, \ldots, x^{(k)}_I$ be the free integer
variables of $\varphi'$ ($k\leq n$). For each assignment of values to these
variables, the subformulas $\phi^{(i)}_I$ are each identically true
or false, hence we have
$$
\varphi \,\equiv\, \bigvee_{(a_1,\ldots,a_k)\in{\ints}^k}
\left(
(x^{(1)}_I, \ldots, x^{(k)}_I) = (a_1,\ldots,a_k)
\,\wedge\,
{\cal B}_{(a_1,\ldots,a_k)}(\phi^{(1)}_F, \ldots, \phi^{(m')}_F)
\right).
$$
Each subformula $\phi^{(i)}_F$ belongs to the theory $\langle {{\reals}},
+, \leq, 1 \nolinebreak \rangle$, which admits the elimination of
quantifiers~\cite{FR79}. The sets of real vectors satisfying these
formulas are thus finite Boolean combinations of linear constraints
with open or closed boundaries. It follows that, for each
$(a_1,\ldots,a_k) \in {\ints}^k$, the set described by ${\cal
B}_{(a_1,\ldots,a_k)}$ is a finite Boolean combination of 
open and closed sets, that is a set belonging to the topological
class ${\cal B}(F)={\cal B}(G)$. Since, according to properties
of the Borel hierarchy, this class forms a subset of $F_\sigma$,
the set described by $\varphi$ is a countable union of countable
unions of closed sets and also lies within
$F_\sigma$.
\end{proof}

%Representing Sets of Integers and Reals with Finite
% Automata
\section{Representing Sets of Integers and Reals with Finite
Automata}
\label{sec-rva}

In this section, we recall the finite-state representation of sets of
real vectors as introduced in~\cite{BBR97}.  

In order to make a finite automaton recognize numbers, one needs to
establish a mapping between these and words. Our encoding scheme
corresponds to the usual notation for reals and relies on an arbitrary
integer base $r > 1$. We encode a number $x$ in base $r$, most
significant digit first, by words of the form $w_I
\star w_F$, where $w_I$ encodes the integer part $x_I$ of $x$ as a finite
word over $\{0, \ldots, r-1\}$, the special symbol ``$\star$'' is a
separator, and $w_F$ encodes the fractional part $x_F$ of $x$ as an
infinite word over $\{0, \ldots, r-1\}$. Negative numbers are
represented by their $r$'s complement. The length $p$ of $|w_I|$,
which we refer to as the {\em integer-part length\/} of $w$, is not
fixed but must be large enough for $-r^{p-1} \leq x_I < r^{p-1}$ to
hold.

According to this scheme, each number has an infinite number of
encodings, since their integer-part length can be increased
unboundedly. In addition, the rational numbers whose denominator
has only prime factors that are also factors of $r$ have two distinct
encodings with the same integer-part length. For example, in base 10,
the number 11/2 has the encodings $005 \star 5(0)^{\omega}$ and $005
\star 4(9)^{\omega}$, ``${\,}^{\omega}$'' denoting infinite
repetition.

To encode a vector of real numbers, we represent each of its components
by words of identical integer-part length. This length can be chosen
arbitrarily, provided that it is sufficient for encoding the vector
component with the highest magnitude. An encoding of a vector
$\vec{x} \in {\reals}^n$ can indifferently be viewed either as a
$n$-tuple of words of identical integer-part length over the alphabet
$\{ 0, \ldots, r-1, \star \}$, or as a single word $w$ over the
alphabet $\{ 0, \ldots, r-1\}^n \,\cup\, \{ \star \}$.

Since a real vector has an infinite number of possible encodings, we
have to choose which of these the automata will recognize. A natural
choice is to accept all encodings. This leads to the following
definition.

\begin{definition}
Let $n > 0$ and $r > 1$ be integers. A {\em Real Vector Automaton (RVA)\/}
$A$ in base $r$ for vectors in ${\reals}^n$ is a B\"uchi automaton
over the alphabet $\{ 0, \ldots, r-1\}^n \,\cup\, \{ \star \}$, such that
\begin{itemize}
\item
every word accepted by $A$ is an encoding in base $r$ of a
vector in ${\reals}^n$, and
\item
for every vector $\vec{x} \in {\reals}^n$, $A$ accepts either
all the encodings of $\vec{x}$ in base $r$, or none of them.
\end{itemize}
\end{definition}

An RVA is said to {\em represent\/} the set of vectors encoded by the
words that belong to its accepted language.

Efficient algorithms have been developed for constructing RVA
representing the sets of solutions of systems of linear equations and
inequations~\cite{BRW98}. Boolean operations can easily be achieved
on RVA by applying the corresponding existing algorithms for
infinite-word automata.

Furthermore, a set represented as an RVA can be quantified
existentially with respect its $i-$th vector component
over the real domain, by replacing each symbol in $\{ 0, \ldots, r-1\}^n$
read by the automaton with the same symbol out of which the $i-$th component
has been removed.  This produces a nondeterministic
automaton that may only accept some encodings of each vector in
the quantified set, but generally not all of them. Such a situation
can arise if the component of highest magnitude for some vectors in the
set is projected out\footnote{For instance, projecting out the first
component of the set $\{(8,1)\}$ in binary would produce an automaton
that does not accept encodings of 1 having less than five bits in their
integer part.}. The second step consists thus of modifying the
automaton so as to make it accept every encoding of each vector that it
recognizes. Algorithms have been developed for this
purpose in the case of finite-word automata~\cite{Boi98,BL01}. These
algorithms also apply to RVA, since the behavior of the underlying
B\"uchi automaton before reading the separator ``$\star$'' is identical
to that of a finite-word automaton recognizing the integer part of
the vectors in the represented set.

Finally, since it is immediate to constrain a number
to be an integer with an RVA by imposing its fractional part to be
either $0^\omega$ or $(r-1)^\omega$ (i.e. by intersecting its
accepted language with $\{0,r-1\}^n\cdot (\{ 0, \ldots, r-1\}^n)^* \cdot
\{\star\} \cdot \{0,r-1\}^n$), it follows that one can construct an
RVA for any formula of the arithmetic theory we are considering.

%Weak Automata and their Properties
\section{Weak Automata and their Properties}
\label{sec-weak}

If one examines the constructions given in~\cite{BRW98} to build
RVA for linear equations and inequations, one notices that they 
have the property that all states within the same
strongly connected component are either accepting or
non accepting. This implies that these automata are {\em weak\/} in the
sense of~\cite{MSS86} (see Section~\ref{sec-background-auto}).

\subsection{Determinizing Weak Automata}

Weak automata have a number of interesting properties. A first one is
that they can be represented both as B\"uchi and co-B\"uchi. Indeed, a
weak automaton $A=(Q,\Sigma, \delta, q_0, F)$ is equivalent
to the co-B\"uchi automaton $A=(Q,\Sigma, \delta, q_0, Q\setminus
F)$, since a run eventually remains within a single component
$Q_i$ in which all states have the same status with respect to being
accepting. A consequence of this is that weak automata can be
determinized by the fairly simple ``breakpoint''
construction~\cite{KV97,MH84} that can be used for co-B\"uchi
automata. This construction is the following. 

Let $A=(Q,\Sigma, \delta, q_0, F)$ be a nondeterministic co-B\"uchi
automaton. The deterministic co-B\"uchi automaton  $A'=(Q',\Sigma,
\delta', q'_0, F')$ defined as follows accepts the same
$\omega$-language~:
\begin{itemize}
\item
$Q' = 2^Q \times 2^Q$, i.e. the states of $A'$ are pairs of sets of states
of $A$.
\item
$q'_0 = (\{q_0\}, \emptyset)$.
\item
For $(S,R) \in Q'$ and $a \in \Sigma$, the transition function is
defined by 
\begin{itemize}
\item
if $R= \emptyset$, then $\delta((S,R),a)= (T,T\setminus F)$ where  $T=
\{q \mid (\exists p \in S)\ q\in\delta(p,a)\}$~: $T$ is
obtained from $S$ as in the classical subset construction, and the
second component of the pair of sets of states is obtained from $T$
by eliminating states in $F$;
\item
if $R \neq \emptyset$, then $\delta((S,R),a)= (T,U\setminus F)$ where  $T=
\{q \mid (\exists p \in S)\ q\in\delta(p,a)\}$, and $U =
\{q \mid (\exists p \in R)\ q\in\delta(p,a)\}$~: the subset
construction set is now applied to both $S$ and $R$ and the states
in $F$ are removed from $U$.
\end{itemize}
\item
$F'= 2^Q \times \{\emptyset\}$.
\end{itemize}
When the automaton $A'$ is in a state $(S,R)$, $R$ represents the
states of $A$ that can be reached by a run that has not gone through
a state in $F$ since the last ``breakpoint'', i.e. state of the form
$(S, \emptyset)$. So, for a given word, $A$ has a run that
does not go infinitely often through a state in $F$ if and only
if $A'$ has a run that does not go infinitely often through a state
in $F'$. Notice that the difficulty that exists for determinizing
B\"uchi automata, which is to make sure that the {\em same}
run repeatedly reaches an accepting state, disappears since,
for co-B\"uchi automata, we are just looking for a run that
eventually avoids accepting states.

It is interesting to notice that the construction implies that all
reachable states $(S,R)$ of $A'$ satisfy $R \subseteq S$. The
breakpoint construction can thus be implemented as a subset
construction in which the states in $R$ are simply tagged, which
implies that the worst-case complexity of the construction is
$2^{{\cal O}(n)}$. This makes the construction behave in practice
very similarly to the traditional subset construction for finite-word
automata.

\subsection{Topological Characterization}

Another property of weak automata that will be of particular interest
to us is the topological characterization of the sets of words that
they can accept. We consider the topology on the sets of infinite words
over an alphabet $\Sigma$ induced by the distance on the $\omega-$words
\[d(w,w')= \left\{\begin{array}{c@{~~~}l}
\frac{1}{|common(w,w')|+1} & \mbox{if $w \neq w'$}\\
0 & \mbox{if $w = w'$},\end{array}
\right.
\]
where $|common(w,w')|$ denotes the length of the longest common
prefix of $w$ and $w'$. The open sets in such a topological space
are the sets of the form $X\cdot \Sigma^\omega$, where 
$X\subseteq \Sigma^+$ is a language of finite words.
Relations between this topology and automata are well understood.
For instance, it has been proved that the languages of
infinite words that can be accepted by a deterministic B\"uchi
automaton are exactly the $\omega-$rational languages
belonging to the class $G_\delta$~\cite{Lan69}.
By duality, deterministic co-B\"uchi automata accept exactly the
$\omega$-regular languages that belong to $F_\sigma$.

As weak deterministic automata can be seen both as deterministic
B\"uchi and deterministic co-B\"uchi, they accept exactly the
$\omega$-regular languages that are in $\fsgd$.
This follows from the results on the Staiger-Wagner class of
automata~\cite{SW74,Stai83}, which coincides with the class of
deterministic weak automata, as can be inferred from~\cite{SW74}
and is shown explicitly in~\cite{MS97}.

\subsection{Inherently Weak Automata}

Given the result proved in Section~\ref{sec-arith}, it is tempting
to conclude that the encodings of sets definable in the theory
$\langle {\reals}, {\ints}, +, \leq \nolinebreak \rangle$ can always
be accepted by weak deterministic automata. This
conclusion is correct, but requires shifting the result from the
topology on numbers to the topology on words, which we will do in the
next section. In the meantime, we need one more result in order to be
able to benefit algorithmically from the fact that we are dealing
with $\fsgd$ sets, i.e. that any deterministic automaton accepting a
$\fsgd$ set is essentially a weak automaton. 

Consider the following definition.
\begin{definition}
A B\"uchi automaton is {\em inherently weak} if none of the reachable
strongly connected components of its transition graph contains both
accepting (including at least one accepting state) and non accepting
(not including any accepting state) cycles.
\end{definition}
Clearly, if an automaton is inherently weak, it can directly be
transformed into a weak automaton~: the partition of the state set is
its partition into strongly connected components and all the states
of a component are made accepting or not, depending on whether the
cycles in that component are accepting or not. 

We will now prove the following.
\begin{theorem}
\label{theo-inh-weak}
Any deterministic B\"uchi automaton that accepts a language in
$\fsgd$ is inherently weak.
\end{theorem}

To prove this, we use the fact that the language accepted by an automaton
that is not inherently weak must have the following property.

\begin{definition}
A language $L \subseteq \Sigma^\omega$ has the {\em dense oscillating
sequence} property if,  $w_1, w_2, w_3, \ldots$ being words and 
$\varepsilon_1, \varepsilon_2, \varepsilon_3, \ldots$ being
distances, one has that $\exists w_1 \forall \varepsilon_1
\exists w_2 \forall \varepsilon_2 \ldots$ such that
$d(w_i,w_{i+1})
\leq \varepsilon_i$ for all $i \geq 1$, $w_i \in L$ for all odd $i$,
and $w_i \not\in L$ for all even $i$.
\end{definition}
Showing that this infinitesimal oscillation is incompatible with the
structure of weak deterministic automata will allow us to conclude.
The proof of Theorem~\ref{theo-inh-weak} can thus be split into the
two following lemmas.

\begin{lemma}
Each $\omega-$language accepted by an B\"uchi automaton that is not
inherently weak has the dense oscillating sequence property.
\end{lemma}

\begin{proof}
Consider a reachable strongly component that contains both an accepting
and a non accepting cycle, and call $p$ a finite word that allows to
reach the first state of the accepting cycle from the initial state
of the automaton. Let $c_A$ (resp. $c_N$)
be the finite word that labels the accepting (resp. non accepting)
cycle, and $t_A$ (resp. $t_N$) a finite word that labels the path
from the first state of the accepting (resp. non accepting) cycle
to the first state of the non accepting (resp. accepting) cycle.

Given an infinite sequence of distances $\varepsilon_1, \varepsilon_2,
\varepsilon_3,\dots$, we are now ready to construct a dense
oscillating sequence for the language $L$ accepted by the automaton.
If $k_2,k_3,k_4,\dots$ is a sequence of natural numbers,
define $u_1=p$, and for all $i>1$~:
$$u_i = \left\{\begin{array}{ll}
u_{i-1}~ c_N^{k_i} ~t_N & \mbox{if}~i~\mbox{is odd} \\
u_{i-1}~ c_A^{k_i} ~t_A & \mbox{if}~i~\mbox{is even.}
\end{array}\right.$$
$w_i$ ($i\geq 1$) is then defined as follows~:
$$w_i = \left\{\begin{array}{ll}
u_i ~ c_A^\omega & \mbox{if}~i~\mbox{is odd} \\
u_i ~ c_N^\omega & \mbox{if}~i~\mbox{is even.}
\end{array}\right.$$
Given $i\geq 1$, it is always possible to find an integer $k_{i+1}$
large enough for \linebreak $d(w_i,w_{i+1})<\varepsilon_i$ to hold.
Indeed, the length of the common prefix between $w_i$ and $w_{i+1}$
increases with $k_{i+1}$. Furthermore, $w_i$ loops either
in an accepting cycle if $i$ is odd, or in a non accepting cycle if
$i$ is even, hence, $w_i\in L$ if and only if $i$ is odd. Thus,
the sequence of $w_i$'s is dense oscillating for the language
accepted by the automaton.
\end{proof}

\begin{lemma}
An $\omega$-regular language that has the dense oscillating sequence
property cannot be accepted by a weak deterministic automaton and
hence is not in $\fsgd$.
\end{lemma}

\begin{proof}
We proceed by contradiction. Assume that a language $L$ having the
dense oscillating sequence property is accepted by a weak
deterministic automaton $A$. Consider the first word $w_1$ in a dense
oscillating sequence for $L$. This word eventually reaches an
accepting component $Q_{i_1}$ of the partition of the state set of
$A$ and will stay within this component. Since $\varepsilon_1$ can be
chosen freely, it can be taken small enough for the run of
$A$ on $w_2$ to also reach the component $Q_{i_1}$ before it starts
to differ from $w_1$. Since $w_2$ is not in $L$, the run of
$A$ on $w_2$ has to eventually leave the component $Q_{i_1}$ and will
eventually reach and stay within a non accepting component $Q_{i_2}
<Q_{i_1}$. Repeating a similar argument, one can conclude that the
run of $A$ on $w_3$ eventually reaches and stays within an
accepting component $Q_{i_3}<Q_{i_2} $. Carrying on with this line of
reasoning, one concludes that the state set of $A$ must contain an
infinite decreasing sequence of distinct components, which is
impossible given that it is finite.
\end{proof}

\subsection{Minimizing Weak Deterministic Automata}

The breakpoint construction reduces much of the determinization
of weak automata to that of finite-word automata. The similarity
can be carried on. Indeed, like finite-word automata, weak
deterministic automata admit a normal form unique up to an
isomorphism~\cite{MS97}.

This normal form can be obtained efficiently using an algorithm proposed
in~\cite{Loe01}. The minimization algorithm consists in locating the strongly
connected components of the graph of the automaton that do not contain
any cycle, then attributing them a new accepting status, according to a
rule involving strongly connected components that are deeper in the graph.
This operation does not affect the language accepted by the automaton,
since for any run $\pi$ of the automaton, $\pi$ cannot loop in such
strongly connected components, leaving $\infi(\pi)$ unchanged. Hopcroft's
classical algorithm for minimizing finite-word automata~\cite{Hop71}
can then be applied directly to the modified weak deterministic
automaton in order to get an equivalent minimal weak deterministic
automaton.

When suitably implemented, this algorithm can be run in time
${\cal O}(n \log n)$, moving us still closer to the
case of automata on finite words.

%Deciding Linear Arithmetic with Real and Integer Variables
\section{Deciding Linear Arithmetic with Real and Integer Variables}

Let us show that the result of Section~\ref{sec-arith} also applies to
the sets of words that encode sets defined in $\langle {\reals},
{\ints}, +, \leq \nolinebreak \rangle$. In order to do so, we need to
establish that the topological class $F_{\sigma} \cap G_{\delta}$
defined over sets of reals is mapped to its $\omega$-word counterpart
by the encoding relation described in Section~\ref{sec-rva}.

\begin{theorem}
Let $n > 0$ and $r > 1$ be integers, and let $L(S) \subseteq (\{ 0,
\ldots, r-1\}^n \cup \{ \star \})^\omega$ be the set of all
the encodings in base $r$ of the vectors belonging to the set $S
\subseteq {\reals}^n$.  If the set $S$ belongs to $\fsgd$ (with respect
to Euclidean distance), then the language $L(S)$ belongs to $\fsgd$ (with
respect to $\omega$-word distance).
\end{theorem}

\begin{proof}
Not all infinite words over the alphabet
$\Sigma=\{ 0, \ldots, r-1 \}^n~\cup~\{ \star \}$ encode a real vector.
Actually, every arbitrary small neighborhood of a word encoding
validly a vector of $\reals^n$ contains words that are not valid
encodings, namely the ones containing multiple occurrences of the
separator ``$\star$'' that are far enough in the word.
Let $V$ be the set of all the valid encodings of vectors in base $r$.
Its complement $\overline{V}$ can be partitioned into a set
$\overline{V}_0$ containing only words in which the separator ``$\star$''
does not appear, and a set $\overline{V}_+$ containing words in which
``$\star$'' occurs at least once (including the words that are not
valid encodings because of an illegal sign digit). Formally, we have
\begin{itemize}
\item $V = \{0,r-1\}^n \cdot (\Sigma \sdiff \{\star\})^* \cdot \{\star\} \cdot
(\Sigma \sdiff \{\star\})^\omega$,
\item $\overline{V}_0 = (\Sigma \sdiff \{\star\})^\omega$,
\item $\overline{V}_+ =  \left(
(\{0,r-1\}^n \cdot \Sigma^* \cdot \{\star\}) \cup
(\Sigma \sdiff \{0,r-1\}^n)
\right)\cdot \Sigma^* \cdot \{\star\} \cdot \Sigma^\omega$.
\end{itemize}

By definition, $V$, $\overline{V}_0$ and $\overline{V}_+$ are disjoint,
and we have $\overline{V}=\overline{V}_0 \cup \overline{V}_+$.
The set $\overline{V}_+$ has the form $X\cdot \Sigma^\omega$ with
$X\subseteq \Sigma^+$, hence it is open.

Similarly, the set $\overline{V}_+ \cup~V$ is open since it can be
expressed as the union of the set $\overline{V}_+$, which 
has just been proved open, and of the set of words beginning by a
valid leading symbol and containing at least one separator, i.e., with
the language $\{0,r-1\}^n\cdot
\Sigma ^* \cdot \{\star\} \cdot \Sigma^\omega$. The latter set is open
for the same reason as $\overline{V}_+$.

Let now consider an open set $S \subseteq {\reals}^n$. 
Each word $w \in L(S)$ has a neighborhood entirely composed of
words in $L(S)$ (formed by the encodings of vectors that belong
to a neighborhood of the vector encoded by $w$) and of words that
contain at least two separators, which
belong to $\overline{V}_+$. Moreover, since $\overline{V}_+$ is open, each
word $w \in \overline{V}_+$ admits a neighborhood fully composed of
words in $\overline{V}_+$. Thus, every word in the language
$L' = L(S)\,\cup\,\overline{V}_+$ has a neighborhood
included in $L'$, implying
that $L'$ is open. Since $L(S) = L' \sdiff \overline{V}_+$, we
have that $L(S)$ is the intersection of an open and a closed set.

The same result holds for a closed set $S \subseteq {\reals}^n$.
Indeed, following the same line of reasoning as above,
$L'=L(\reals^n \sdiff S) \cup \overline{V}_+$ is open
because the complement of $S$ is itself open. On the other hand, 
we have $L(\reals^n \sdiff S)=\overline{L(S)}\:\cap\:V$.
Therefore, $\overline{L'}= L(S) \cup \overline{V}_0$ holds,
hence $L(S) = \overline{L'} \sdiff \overline{V}_0 =
\overline{L'}\:\cap\:(\overline{V}_+ \cup V)$.
The last relation entails that
$L(S)$ is the intersection of a closed and an open set.

We are now ready to address the case of a set $S \subseteq {\reals}^n$
that belongs to $\fsgd$. Since $S$ is in $F_{\sigma}$, it can be
expressed as a countable union of closed sets $S_1,S_2, \dots$~.
It has been showed that the languages $L(S_1), L(S_2), \dots$
are Boolean combinations of open and of closed sets, and thus belong
to the topological class $F_{\sigma}$. Therefore, $L(S) = L(S_1)
\cup L(S_2) \cup \cdots$ is a countable union of sets in $F_{\sigma}$,
and belongs itself to $F_{\sigma}$. Now, since $S$ is in $G_{\delta}$,
it can also be expressed as a countable intersection of open sets
$S'_1, S'_2, \dots$~. The languages $L(S'_1), L(S'_2), \dots$
belong to the topological class $G_{\delta}$. Hence, $L(S) = L(S'_1)
\cap L(S'_2) \cap \cdots$ is a countable intersection of sets in $G_{\delta}$,
and thus belongs itself to $G_{\delta}$. Therefore, we have
$L(S) \in \fsgd$.
\end{proof}

Knowing that the language of the encodings of any set definable in the
theory $\langle {\reals}, {\ints}, +,$ $\leq \nolinebreak \rangle$
belongs to $\fsgd$, we use the results of Section~\ref{sec-weak} to
conclude the following.

\begin{theorem}
\label{theo-rva-weak}
Every deterministic RVA representing a set definable in $\langle
{\reals}, {\ints}, +,$ $\leq \nolinebreak \rangle$ is inherently weak.
\end{theorem}

This property has the important consequence that the construction and
the manipulation of RVA obtained from arithmetic formulas can be
performed effectively by algorithms operating on weak deterministic
automata.  Precisely, to obtain an RVA for an arithmetic formula one
can proceed as follows. 

For equations and inequations, one uses the constructions given
in~\cite{BRW98} to build weak RVA.  Computing the intersection,
union, and Cartesian product of sets represented by RVA simply
reduces to performing similar operations with the languages accepted
by the underlying automata, which can be done by simple product
constructions. These operations preserve the weak nature of the
automata.  To complement a weak RVA, one determinizes it using the
breakpoint construction, which is guaranteed to yield an inherently
weak automaton (Theorem~\ref{theo-rva-weak}) that is easily converted
to a weak one. This deterministic weak RVA is then complemented by
inverting the accepting or non-accepting status of each of its
components, and then removing from its accepted language the words
that do not encode validly a vector (which is done by means of an
intersection operation).

An existential quantifier can be applied to a set represented as an
RVA by using the construction detailed in Section~\ref{sec-rva}.
This operation does not affect the weak nature of the automaton,
which can then be determinized by the breakpoint construction. The
determinization algorithm has to produce an inherently weak RVA
easily converted to a weak automaton. 

Thus, in order to decide whether a formula of $\langle {\reals}, {\ints},
 +,$ $\leq \nolinebreak \rangle$ is satisfiable, one simply builds an RVA
representing its set of solutions, and then check whether this
automaton accepts a nonempty language. This also makes it possible to
check the inclusion or the equivalence of sets represented by
RVA. The main result of this paper is that, at every point of the
interpretation of a formula, the constructed automaton remains weak
and thus only the simple breakpoint construction is needed as a
determinization procedure.

Finally, as weak deterministic automata can be efficiently
minimized, each constructed automaton can be reduced down to a normal
form. This is particularly useful from a practical point of view,
since it speeds up the comparisons between sets by reducing them
to structural tests on the automata, and since it prevents the
representations from becoming unnecessarily large.

%Implementation
\section{Experiments}

The decision procedure proposed in this paper has been implemented
successfully in the LASH toolset, a package based on finite-state
automata for representing infinite sets and exploring infinite state
spaces~\cite{Boi99}.

\begin{figure}
\begin{center}
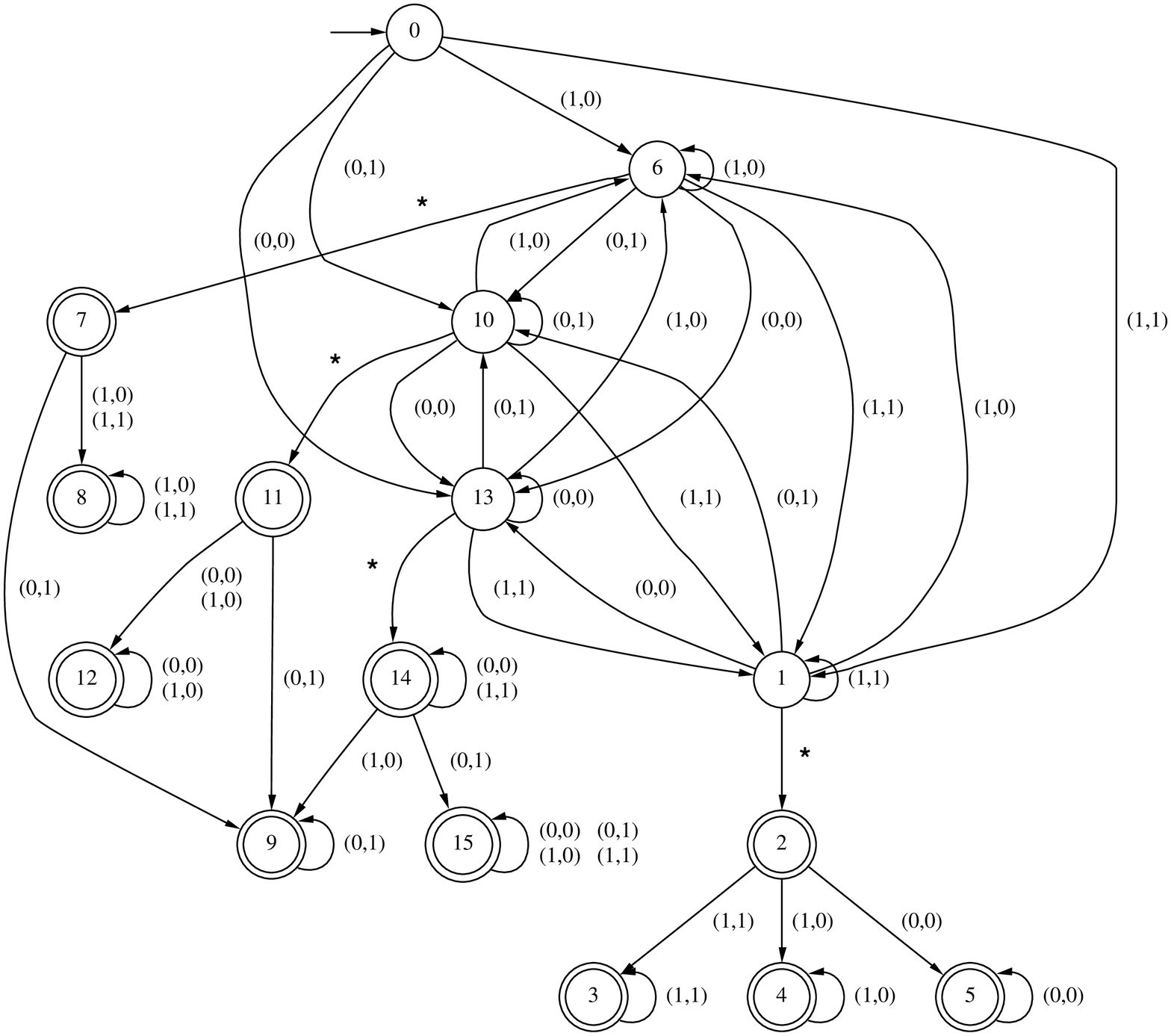
\end{center}
\caption{Periodic tiling with triangles.~\label{fig-triangles}}
\end{figure}

\begin{figure}
\begin{center}
\includegraphics[width=11cm]{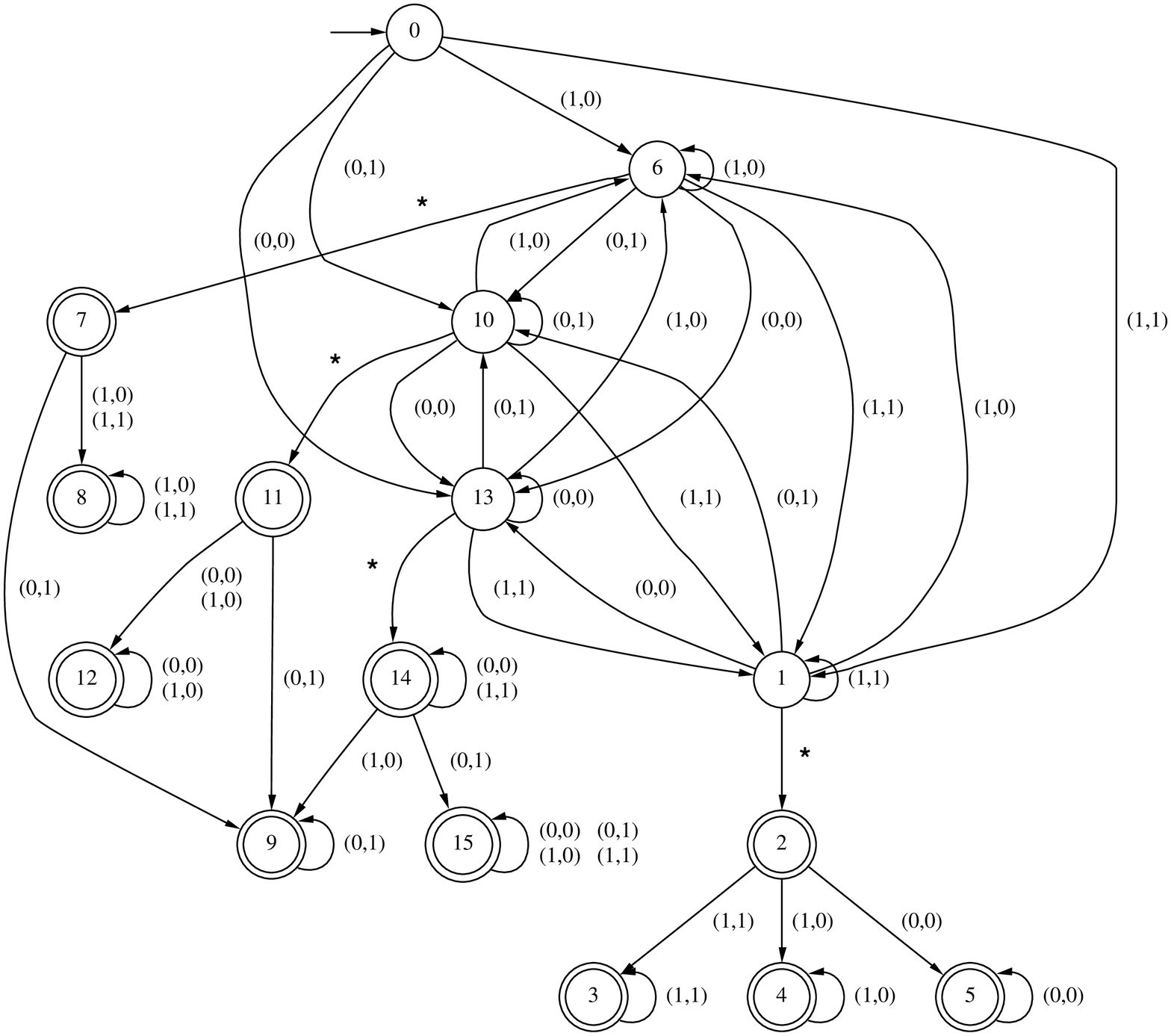}
\end{center}
\caption{Weak RVA representing the periodic tiling in 
	binary.~\label{fig-triangles-automaton}}
\end{figure}

Various experiments have been achieved with the RVA package.
For instance, it is possible to represent the set of
Figure~\ref{fig-triangles}, which combines discrete
and continuous features, by a weak RVA. Indeed, this set is defined by
the following formula of the additive theory over the reals and integers~:
$$\begin{array}{l}
\{(x_1,x_2)\in \reals^2\mid 
(\exists x_3,x_4\in \reals) (\exists x_5,x_6\in \ints)\\
~\hspace{1cm}(x_1=x_3+2 x_5 \wedge
x_2=x_4+2 x_6 \wedge
x_3\geq 0\wedge x4\leq 1\wedge x_4\geq x_3)\}.
\end{array}$$
This set admits the compact minimal representation of
Figure~\ref{fig-triangles-automaton}.

\begin{figure}
\begin{center}
\includegraphics[width=11.8cm]{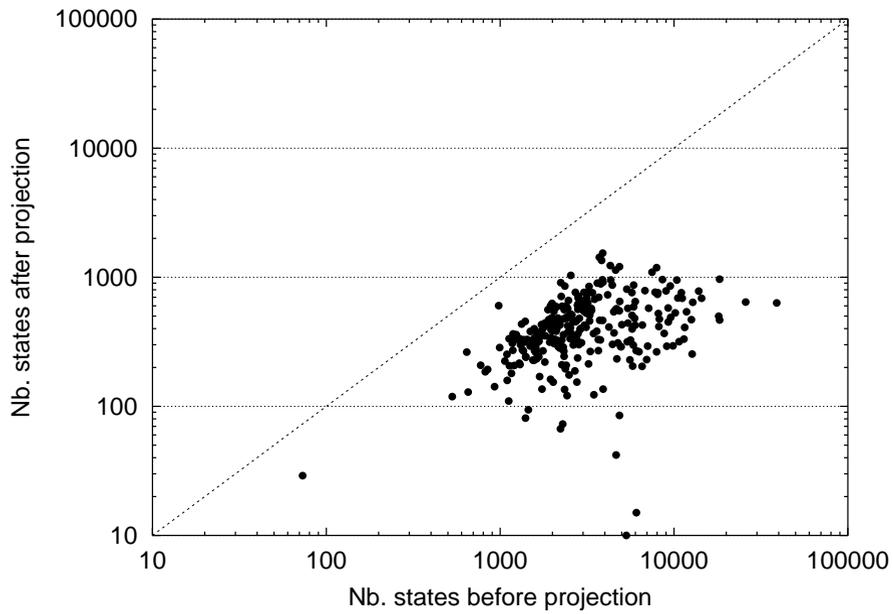}
\end{center}
\caption{The effect of projection-determinization
	on RVA.\label{fig-rva}}
\end{figure}

One might fear that the exponential worst-case complexity of the
breakpoint determinization algorithm makes our decision 
procedure unusable. Experimental results however show that such a
blow-up does not frequently occur in practical applications. As an
illustration, Figure~\ref{fig-rva} shows the cost of
projecting and then determinizing the finite-state representations
of some periodic subsets of $\reals^3$ obtained by combining linear
constraints with arbitrary coefficients, and then by inducing a
periodicity by means of an integer quantification. The interesting
observation is that the finite-state representations have
always less states after the projection than before, whereas an
exponential blow-up could have been feared.

\begin{figure}
\begin{center}
\includegraphics[width=11.8cm]{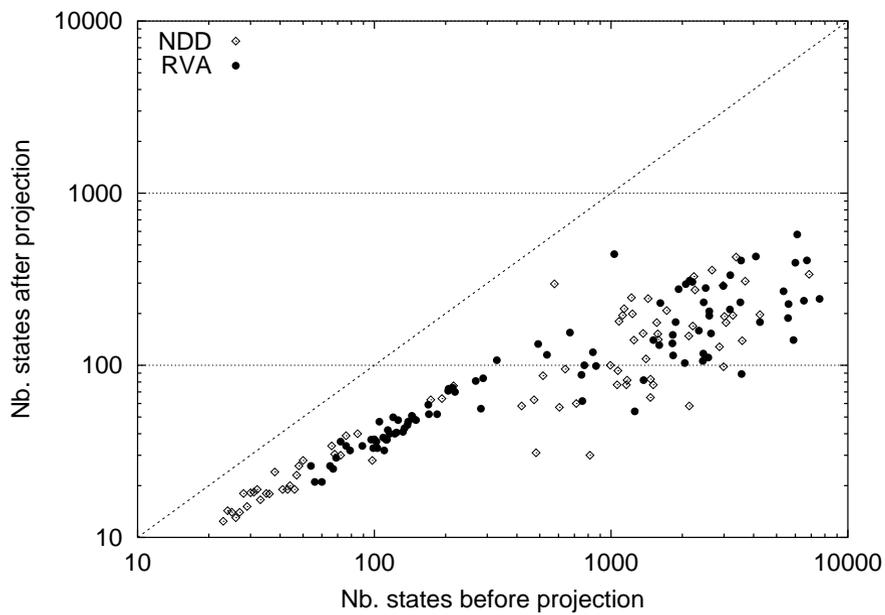}
\end{center}
\caption{The effect of projection-determinization on
	NDD and RVA.\label{fig-rva-ndd}}
\end{figure}

Another finite-state representation system, the NDD ({\em Number
Decision Diagram\/})~\cite{WB95,Boi98}, is based on finite-word
automata and is able to represent the
subsets of $\ints^n$ that can be expressed in an extension of the
first-order theory $\langle {\ints}, +, \leq \nolinebreak \rangle$.
Figure~\ref{fig-rva-ndd} compares the size of weak RVA with that of
NDD representing the same subsets of $\ints^3$ obtained by combining
linear constraints with arbitrary coefficients. One notices that
the behavior of RVA is very similar to that of NDD, that are reputed
to behave quite well in practice~\cite{WB00}.

These observations make one think that the pathological conditions
that lead the breakpoint construction to blow-up are seldom met in
practice.

%Conclusions
\section{Conclusions}

A probably unusual aspect of this paper is that it does not introduce
new algorithms, but rather shows that existing algorithms can be used
in a situation where {\em a priori\/} they could not be expected to
operate correctly. To put it in other words, the contribution is not
the algorithm but the proof of its correctness. 

The critical reader might be wondering if all this is really
necessary. After all, algorithms for complementing B\"uchi automata
exist, either through determinization~\cite{Saf88} or
directly~\cite{Buc62,SVW87,KV97,Kla91} and the more recent of these
are even fairly simple and potentially implementable. There are no
perfectly objective grounds on which to evaluate ``simplicity'' and
``ease of implementation'', but it is not difficult to convince
oneself that the breakpoint construction for determinizing weak automata
is simpler than anything proposed for determinizing or complementing
B\"uchi automata. Indeed, it is but one step of the probably simplest
complementation procedure proposed so far, that of~\cite{KV97}.
Furthermore, there is a complexity improvement from
$2^{O(n \log n)}$ to $2^{O(n)}$, and being able to work with deterministic
weak automata allows minimization~\cite{Loe01}, which leads to a
normal form. Those claims to simplicity and ease of implementation
are substantiated by the experimental results.

Our implementation makes it possible to represent possibly non convex
periodic sets containing both integers and reals, and to manipulate
those sets using Boolean operations and quantification, and to check
relations existing between them.  To the best of our knowledge, doing
this is beyond the scope of any other implemented tool. The potential
application field of RVA is wide and range from symbolic analysis of
linear hybrid systems~\cite{ACHHHNOSY95} to temporal
databases~\cite{CI88,KSW90}.

\bibliography{paper}
\bibliographystyle{acmtrans}

\end{document}